\documentstyle[12pt]{article}
\newcommand{\beq}{\begin{equation}}
\newcommand{\eeq}{\end{equation}}
\newcommand{\beqa}{\begin{eqnarray}}
\newcommand{\eeqa}{\end{eqnarray}}
\newcommand{\n}{\nonumber\\}
\renewcommand{\theequation}{\mbox{\arabic{section}.\arabic{equation}}}
\begin{document}
\baselineskip 6mm
\begin{titlepage}
\hfill YITP/U-92-27 \\
\vskip -5mm
\hfill August, 1992
\\
\vskip 1cm

\begin{center}
{\large\bf Rational Conformal
Field Theory and
Multi-Wormhole Partition Function
in 3-dimensional Gravity\\}

\vskip 15mm

Shun'ya Mizoguchi${}^{\dag}$\\

\vskip 10mm

{\it Uji Research Center \\
Yukawa Institute for Theoretical Physics \\
Kyoto University, Uji 611, Japan \\}
\end{center}
\vskip 3cm
\noindent
{\bf Abstract.}\hskip 5mm
We study the Turaev-Viro invariant as the Euclidean Chern-Simons-Witten
gravity partition function with positive cosmological constant.
After explaining why it can be identified as the partition function of
3-dimensional gravity, we show that the initial data of the TV invariant
can be constructed from the duality data of a certain class of rational
conformal field theories, and that, in particular, the original
Turaev-Viro's initial data is associated with the $A_{k+1}$ modular
invariant WZW model. As a corollary we then show that
the partition function $Z(M)$ is bounded from above
by $Z((S^2\times S^1)^{\sharp g})
=(S_{00})^{-2g+2}\sim \Lambda^{-\frac{3g-3}{2}}$,
where $g$ is the smallest genus of handlebodies with which $M$
can be presented by Hegaard splitting.
$Z(M)$ is generically very large near $\Lambda\sim +0$
if $M$ is neither $S^3$ nor a lens
space, and many-wormhole
configurations dominate near $\Lambda\sim +0$ in the sense
that $Z(M)$ generically tends to
diverge faster as the ``number of
wormholes'' $g$ becomes larger.

\vfill\footnoterule
{${}^{\dag}$\footnotesize Yukawa fellow. e-mail: mizo@jpnyitp.bitnet}
\end{titlepage}

\renewcommand{\Large}{\normalsize}
\renewcommand{\large}{\normalsize}

\section{Introduction}

\indent

	Recently it has been shown that the Turaev-Viro (TV)
topological invariant constructed as a state-sum over $q$-deformed
spin-networks \cite{TV} provides us an interesting 3-dimensional
Euclidean gravity model [2-5].
 A crucial observation is
that their construction strongly resembles the 3-dimensional Regge
calculus studied by Ponzano and Regge \cite{PR} in the late 60's.
In ref.\cite{PR} they assigned an SU(2) $6j$ symbol to each simplex
of a triangulated 3-manifold,
and then summed them up over all spin-configurations
with some edge and vertex factors, regarding spins as actual lengths
of edges. They found that the summation can be written as a path-integral
of 3-dimensional gravity action in the large spin region.
Surprisingly, the TV invariant, constructed after more than twenty years,
was completely in the same form as Ponzano-Regge's partition function,
except that the quantities associated with representations of SU(2)
were replaced by those of $q$-deformed SU(2)
with a root of unity $q$.
In fact the TV invariant is, as explained later, identified with
the partition function of the Euclidean Chern-Simons-Witten (CSW)
gravity theory with positive cosmological constant \cite{CSW}.\footnote{
It should be noted that by CSW partition function we mean
{\it pure} Chern-Simons partition function without factoring out the global
diffeomorphisms \cite{CD}. See below.
}
We first show that the TV invariant can be constructed from duality data
of a certain class of rational conformal field theories (RCFT's),
and that, in particular, the original Turaev-Viro's initial data
which is relevant to 3-dimensional gravity is associated with
the $A_{k+1}$ modular invariant SU(2) WZW model. As a corollary
we then show that
the partition function $Z(M)$ is bounded from above
by $Z((S^2\times S^1)^{\sharp g})=(S_{00})^{-2g+2}
\sim \Lambda^{-\frac{3g-3}{2}}$,
where $g$ is the smallest genus of handlebodies with which $M$
can be presented by Hegaard splitting.
This upper-bound argument is motivated by the work by Kohno \cite{Kohno},
in which Witten's topological invariant (Jones polynomial
as an expectation value of a Wilson line which carries
identity representation)\cite{WitJones}
is studied as a Hegaard splitting invariant, though in that context
the estimation is for the lower-bound of $g$.

	Several years ago a possible scenario that the wormhole
sum may force the cosmological constant to be zero was proposed and
discussed \cite{wormhole}. This idea has been also tested in
$2+1$-dimensional CSW gravity by summing up $S^2\times S^1$
wormholes to obtain a rather negative result \cite{CD}.
In this topological theory the partition function does not
depend on size or location of wormholes, and the wormhole summation
is reduced to a more tractable problem than in 4 dimensions.
However, we still do not know a natural weight with which we sum
over topologies since we do not know a 3-dimensional analogue of string
field theory or matrix model, so it would be meaningful at this stage
to study the cosmological constant dependence of a partition function
with fixed topology.
In general one can probe the large $k$ behavior of
the Chern-Simons (CS) partition function by the saddle-point
approximation \cite{CD,WitJones,WitTopChange,C,C2}, in which
a Gaussian integration around each extremum gives the Ray-Singer (RS)
torsion multiplied by some phase factor.
The full ${\rm SU(2)}\times{\rm SU(2)}$
CS partition function,
which is considered as an Euclidean counterpart of the CSW
gravity with positive cosmological constant (without exotic terms),
is given by the absolute square of the sum (or the integration
over the moduli) of such contributions.
Instead, once one recognizes the correspondence
between the TV invariant and the CS theory,
one may use it to determine the exact cosmological constant
behavior away from $\Lambda\sim (+)0$.
We thus expect the TV invariant to be an alternative
to the saddle-point approximation in 3-dimensional gravity.

	This paper is organized as follows. In sect.2 we briefly
review the construction of the TV invariant, and then explain how
it is related to the CSW gravity theory. In sect.3 we introduce
the axiom of RCFT \cite{MS} and show that
the TV invariant can be constructed from a certain class of RCFT's.
In sect.4 we study the TV invariant as multi-wormhole partition functions,
and investigate their topology and cosmological constant dependence,
in particular their behavior near $\Lambda\sim+0$.
Sect.5 contains a brief summary of our results and a discussion.
In appendix we give an elementary
proof of the factorization formula of the TV invariant,
which is used in sect.4.

\section{3-dimensional gravity and the TV Invariant}
\subsection{Review of the TV invariant}
\setcounter{equation}{0}

\indent

	We will begin by the definitions first.
Let $I$ be a finite set of ``spin'' variables.
Assume that we have distinguished a set $adm$ of unordered triples of elements
of $I$. The triple $(i,j,k)$ is said to be admissible
if $(i,j,k)\in adm$. An ordered 6-tuple $(i,j,k,l,m,n)\in I^6$
is called admissible if the unordered $(i,j,k)$, $(k,l,m)$,
$(m,n,i)$ and $(j,l,n)$ are admissible. Next assume that we are given
a complex-valued function
$\left|\begin{array}{ccc}i&j&k\\l&m&n\end{array}\right|$,
which is called symbol, of admissible 6-tuple
$(i,j,k,l,m,n)$. It is assumed to have the following symmetries:
\beq
\left|\begin{array}{ccc}i&j&k\\l&m&n\end{array}\right|=
\left|\begin{array}{ccc}j&i&k\\m&l&n\end{array}\right|=
\left|\begin{array}{ccc}i&k&j\\l&n&m\end{array}\right|=
\left|\begin{array}{ccc}i&m&n\\l&k&j\end{array}\right|=
\left|\begin{array}{ccc}l&m&k\\i&j&n\end{array}\right|\;,
\label{eq:sym}
\eeq
so that it is naturally associated with a tetrahedron.
A 6-tuple is admissible if and only if
any of the four triples that forms a triangle of the associated tetrahedron
is admissible.
Finally we assume that we are given a complex-valued
function $w(i)\equiv w_i$ of $I$, and a complex number $w(\neq 0)$.
The symbol
$\bigl| \mathop{}^{\cdots}_{\cdots} \bigr|$,
the functions $w_i$ and the number $w$
are collectively referred to as initial data.

	By a colored tetrahedron
we mean a tetrahedron with an element of $I$
attached to each edge. By a coloring ${\phi}$ of edges
$\{E_1,\ldots,E_b\}$
we mean a mapping
$\phi:\{E_1,\ldots,E_b\}\rightarrow I$; in other word, the way how we assign
spins to the edges.

	After these definitions
Turaev-Viro's theorem is stated
as follows \cite{TV}. Let $M$ be a compact triangulated 3-manifold.
Let $a$ be the number of vertices, $e$ of which lie on the boundary
$\partial M$. Let $E_1,\ldots, E_b$ be the edges of $M$,
the first $f$ of which belong to $\partial M$. Finally let $T_1,\ldots,T_d$
be the tetrahedra of $M$.
Turaev and Viro introduced the following three conditions on the initial data
\cite{TV}:
\beqa
(*)&&\sum_j w_j^2w_{j_4}^2
\left|\begin{array}{ccc}j_2&j_1&j\\j_3&j_5&j_4\end{array}\right|
\left|\begin{array}{ccc}j_3&j_1&j_6\\j_2&j_5&j\end{array}\right|
=\delta_{j_4,j_6}\n
(**)&&\sum_j w_j^2
\left|\begin{array}{ccc}j_2&a&j\\j_1&c&b\end{array}\right|
\left|\begin{array}{ccc}j_3&j&e\\j_1&f&c\end{array}\right|
\left|\begin{array}{ccc}j_3&j_2&j_{23}\\a&e&j\end{array}\right|\n
&&=
\left|\begin{array}{ccc}j_{23}&a&e\\j_1&f&b\end{array}\right|
\left|\begin{array}{ccc}j_3&j_2&j_{23}\\b&f&c\end{array}\right|\n
(***)&&
w^2=w_j^{-2}\sum_{\scriptsize\begin{array}{c}
k,l\\(j,k,l)\in adm\end{array}}
w_k^2w_l^2\;\;\; \mbox{\rm for all $j\in I$}.
\label{eq:initial data}
\eeqa
Here the sum in $(*)$ and $(**)$ are carried out over $j$ such that
all the symbols involved in them are defined.
If the initial data satisfy these three conditions
$(*)$,$(**)$ and $(***)$, then the quantity $\Omega_M(\alpha)$ such that
\beq
\Omega_M(\alpha)=\sum_{\phi\in adm(M,\alpha)}|M|_{\phi} \label{eq:Omega_M}
\eeq
\beq
|M|_{\phi}=w^{-2a+e}
\prod_{r=1}^f w_{\phi(E_r)}
\prod_{s=f+1}^b w^2_{\phi(E_s)}
\prod_{t=1}^d |T^{\phi}_t|
\eeq
is independent of the triangulation of $M$, but depends only on the topology
of $M$ and the coloring of $\partial M$. Here $adm(M,\alpha)$ is the set
of all admissible colorings with the fixed coloring $\alpha$ of $\partial M$,
and $|T^{\phi}_t|$ stands for the symbol associated to the $t$th tetrahedron.

	The outline of the proof is the following. Due to the Alexander's
theorem any two triangulated 3-manifolds with the same topology can be
transformed each other by some sequence of the Alexander moves
and the inverse transformations. If one passes
from the triangulation to the dual cell subdivision (fig.1), an Alexander move
can be generated by compositions of three kinds of elementary moves,
{\it i.e.} the bubble move ${\cal B}$, the lune move ${\cal L}$
and the Matveev move ${\cal M}$ (fig.2). They first translated $\Omega_M$
in terms of the dual cell subdivision, and then showed that
it is invariant under ${\cal L}$, ${\cal M}$ and ${\cal B}$
if the initial data satisfy the condition $(*)$, $(**)$ and $(***)$,
respectively.

	For future convenience we define \cite{KMS}
\beq
c(l,k)= w_k^{-2}w_l^{-2}\sum_j\delta (l,j,k) w_j^2
\label{eq:c(l,k)}
\eeq
and
\beq
\tilde{w}^2=\sum_{i\in I}w_i^4,
\eeq
where $\delta (l,j,k)=1$ if $(l,j,k)$ is admissible, 0 otherwise.
Then
\beqa
\sum_{i,j}w_i^2w_j^2
\left|\begin{array}{ccc}i&j&k\\l&m&k\end{array}\right|^2
&=&
\sum_jw_i^{-2}w_j^{-2}\delta (l,m,k)\delta (l,j,k)\n
&=&
w_k^2c(l,k)\delta (l,m,k),
\eeqa
or another calculation gives
\beqa
\phantom{
\left|\begin{array}{ccc}i&j&k\\l&m&k\end{array}\right|^2
}
&=&
w_k^2c(m,k)\delta (l,m,k),
\eeqa
so $c(l,k)=c(m,k)$ if $(l,m,k)$ is admissible.
$I$ is said to be irreducible
if for any $j,m\in I$
there exists a sequence $l_1,\ldots,l_n$ with $l_1=j$, $l_n=m$
such that $(l_i,l_{i+1},l_{i+2})$ is admissible for any $i=1,\ldots,n-2$.
Obviously $c(i,j)$ is a constant if $I$ is irreducible \cite{TV,KMS}.
In that case
\beqa
(w^2&=)&w_j^{-2}\sum_{k,l}w_k^2w_l^2\delta (j,k,l)\n
&=&\sum_kw_k^4\cdot w_j^{-2}w_k^{-2}\sum_lw_l^2\delta (j,k,l)\n
&=&\tilde{w}^2c\;\;\;(c=c(j,k)),
\eeqa
and so $(***)$ is satisfied. This fact will be used in the next section.

\subsection{Identification with the CSW gravity partition function}

\indent

	For the following reason
$\Omega_M(\alpha)$ can be identified as the partition function of
3-dimensional quantum gravity \cite{MT}.
Let us evaluate $\Omega_M(\alpha)$
on a tetrahedron as a triangulated 3-ball $B^3$:
\beq
\Omega_{B^3}(\alpha)=w^{-2a+4}\prod_{r=1}^6w_{\alpha(E_r)}\sum_{\phi(\alpha)}
\prod_{s=7}^{b}w_s^2\prod_{t=1}^d|T_t^{\phi}|. \label{eq:3ball}
\eeq
Here $\alpha$ represents a fixed coloring of the boundary.
We fix a root of unity of degree $2(k+2)$, $q$, such that
$q^2$ is a primitive root of unity of degree $k+2$.
As initial data we take\footnote{$q_0$ in ref.\cite{TV}
corresponds to $q$ we use in this paper.}
\begin{eqnarray}
&&I=\{0,\frac{1}{2},1,\ldots,\frac{k}{2}\} ~~(k\in\mbox{\rm {\bf Z}})\n
&&|T_i^{\phi}|=(-1)^{\sum_{l=1}^6 j^{(i)}_l}
\left\{
\begin{array}{ccc}
j^{(i)}_1&j^{(i)}_2&j^{(i)}_3 \\ j^{(i)}_4&j^{(i)}_5&j^{(i)}_6
\end{array}
\right\} \n
&&w_j^2=(-1)^{2j}[2j+1],\hskip 5mm w^2=-\frac{2(k+2)}{(q-q^{-1})^2},
\label{eq:su(2) initial data}
\end{eqnarray}
where
$\bigl\{ \mathop{}^{\cdots}_{\cdots} \bigr\}$ stands for
the Racah-Wigner
${\cal U}_q{\rm su}(2)$ $q$-$6j$ symbol\cite{KR}:
\beqa
&&\left\{
\begin{array}{ccc}
j_1&j_2&j_{12} \\ j_3&j&j_{23}
\end{array}
\right\}\n
&=&\Delta(j_1,j_1,j_{12})\Delta(j_3,j,j_{12})
\Delta(j_1,j,j_{23})\Delta(j_3,j_2,j_{23})\sum_{z\geq 0}(-1)^z[z+1]!\n
&&\cdot([z-j_1-j_2-j_{12}]![z-j_3-j-j_{12}]!
[z-j_1-j-j_{23}]![z-j_3-j_2-j_{23}]!\n
&&\;\cdot [j_1+j_2+j_3+j-z]![j_1+j_3+j_{12}+j_{23}-z]!
[j_2+j+j_{12}+j_{23}-z]!)^{-1}\n
\eeqa
with
\beq
\Delta(a,b,c)=\sqrt{
\frac{[-a+b+c]![a-b+c]![a+b-c]!}
{[a+b+c+1]!}
}
\eeq
and $[n]=\frac{q^n-q^{-n}}{q-q^{-1}}$.
A triple $(i,j,l)\in I^3$ is defined to be admissible
if $i+j+l\in\mbox{\rm {\bf Z}}$, $\leq k+2$ and
$i\leq j+l,~~j\leq l+i,~~l\leq i+j$.

	Suppose that $k$ is very large. Then up to $O(k^{-2})$
a $q$-$6j$ symbol above becomes an ordinary Racah-Wigner $6j$ symbol of su(2).
In the semi-classical continuum limit \cite{PR,HP}(values of spin
$\phi(E_s)\rightarrow\infty$, number of vertices $a\rightarrow\infty$)
a $6j$ symbol behaves as \cite{SG}
\begin{equation}
\left\{
\begin{array}{ccc}
j_1&j_2&j_3 \\ j_4&j_5&j_6
\end{array}
\right\}
\approx
(\frac{1}{12\pi V})^{\frac{1}{2}}
\cos(\sum_{i=1}^6\theta_iJ_i+\frac{\pi}{4}) \label{eq:asympt}
\end{equation}
in the domain where $J_i$ are uniformly large ($J=j+\frac{1}{2}$).
Here $V$ is the volume of the tetrahedron, and $\theta_i$ is
the angle between the outer normal of the two faces which have the edge $j_i$
in common, regarding $J$ as actual length.
Replacing $q$-$6j$ symbols in (\ref{eq:3ball})
by (\ref{eq:asympt}), we obtain

\beqa
\Omega_M(\alpha)&=&
\left(\frac{k^3}{2\pi}\right)^{-2a+4}
\prod_{r=1}^6 w_{\alpha(E_r)}
\n
&&\cdot\sum_{\phi}
\prod_{s=7}^{b}(-1)^{2\phi(E_s)}(2\phi(E_s)+1)
\prod_{t=1}^d (-1)^{\sum_{l=1}^6 j_l^{(t)}}
\left\{
\begin{array}{ccc}
j_1^{(t)}&j_2^{(t)}&j_3^{(t)} \\ j_4^{(t)}&j_5^{(t)}&j_6^{(t)}
\end{array}
\right\}
\n
&\approx
&
\left(\frac{k^3}{2\pi}\right)^{-2a+4}
\prod_{r=1}^6 w_{\alpha(E_r)}
\sum_{\phi}
\prod_{s=7}^{b}(-1)^{2\phi(E_s)}(2\phi(E_s)+1)
\n
&&\cdot\prod_{t=1}^d (-1)^{\sum_{l=1}^6 j_l^{(t)}}
(\frac{1}{12\pi V^{(t)}})^{\frac{1}{2}}
\cos(\sum_{l=1}^6\theta_l^{(t)}(j_l^{(t)}+\frac{1}{2})+\frac{\pi}{4}),
\eeqa
where $j_l^{(t)}=\phi(E_s)$ if the $l$th edge of the $t$th tetrahedron
is $E_s$.
If we take {\it only the positive frequency part} of cosine,
we have
\beqa
&&\Omega_M(\alpha)~(\mbox{\rm positive frequency part})\n
&\approx&
\left(\frac{k^3}{2\pi}\right)^{-2a+4}
\left(\frac{i}{48\pi}\right)^{\frac{d}{2}}
\prod_{r=1}^6 w_{\alpha(E_r)}
\n
&&\cdot
\sum_{\phi}
\prod_{s=7}^{b}(-1)^{2\phi(E_s)}(2\phi(E_s)+1)
\prod_{t=1}^d \frac{1}{\sqrt{V^{(t)}}}
\exp\{
i\sum_{l=1}^6(\pi-\theta_l^{(t)})(j_l^{(t)}+\frac{1}{2})
\}
\n
&=&
\left(\frac{k^3}{2\pi}\right)^{-2a+4}
\left(\frac{i}{48\pi}\right)^{\frac{d}{2}}
(\prod_{r=1}^6 w_{\alpha(E_r)})^{-1}
\n
&&\cdot
\sum_{\phi}
\prod_{s=1}^{b}
\left[
\frac{(-1)^{2\phi(E_s)}(2\phi(E_s)+1)}
{(\prod_{j=1}^{n_s}V^{(s,j)})^{\frac{1}{12}}}
\exp
\{i\sum_{j=1}^{n_{s}}(\pi-\theta^{(s,j)})(\phi(E_s)+\frac{1}{2})\}
\right].
\label{eq:PFP}
\eeqa
In the last line we changed the summation over all tetrahedra
to the double summation;
first over $n_s$ tetrahedra which have the edge $E_s$ in common,
and then over all edges.

	If the triangulation $\phi$ were such that the manifold
could be embedded in a flat Euclidean geometry,
the sum $\sum_{j=1}^{n_{s}}(\pi-\theta^{(s,j)})$ should be $2\pi$.
In our case, however, $\phi$-summation is carried out over colorings
that generically can not be embedded in a flat geometry, so we may regard
$\phi$-summation as integration over metric. More precisely,
the summation $\sum_{s=1}^b\sum_{j=1}^{n_s}
(\pi-\theta^{(s,j)})(\phi(E_s)-\frac{1}{2})$ can be considered
as a realization of the Einstein-Hilbert action on a 3-dimensional
simplicial decomposition
\cite{PR}. To show this, suppose that we parallel-transport some tangent
vector along a small loop.
If the loop encloses no edges,
the vector does not change after the parallel-transport. If, however,
the loop encloses an edge $E_s$, the vector rotates by angle
$\sum_{j=1}^{n_s}(\pi-\theta^{(s,j)})-2\pi$. Hence the curvature tensor has
its support only at each edge. Integrating the scalar curvature $R$
over the interior of a thin cylinder $C$ along $E_s$, we obtain
\beqa
\int_Cd^3x R&=&(\phi(E_s)+\frac{1}{2})\cdot\int_{{\mbox{\rm\scriptsize
section of}}\; C}d^2x R\n
&=&
(\sum_{j=1}^{n_s}(\pi-\theta^{(s,j)})-2\pi)(\phi(E_s)+\frac{1}{2}).
\label{eq:deficitangle}
\eeqa
Substituting (\ref{eq:deficitangle}) into (\ref{eq:PFP})
gives
\beq
(\mbox{\rm const. which depends only on the boundary})
\times\int{\cal D}\phi\exp i\int_M\sqrt{g}R,
\label{eq:action}
\eeq
where
\beq
{\cal D}\phi=\sum_{\phi}\prod_{s=1}^b
\frac{2\phi(E_s)+1}{(\prod_{j=1}^{n_s}V^{(s,j)})^{\frac{1}{12}}}.
\label{eq:dphi}
\eeq
Therefore the positive frequency part of $\Omega_M(\alpha)$
can be seen as a partition function of 3-dimensional gravity with measure
${\cal D}\phi$ in the large $k$ and the semi-classical continuum limit.
This interesting suggestion on the relation between spin net-works
and  quantum gravity was made by Ponzano and Regge in the late 60's \cite{PR}.
In fact, since they deal with classical su(2) symbols, the expression
for the partition function diverges. On the other hand, due to the restriction
for the spin variables in $q$-$6j$ symbols \cite{KR}
the TV invariant is finite
and well-defined. Recently the next-leading term in the action
has been estimated and found to be
a cosmological term with cosmological constant
$\Lambda=\frac{4\pi^2}{k^2}+O(k^{-4})$ in this approximation \cite{MT}.

	Although it is interesting, some subtleties had been remaining
unsolved until recently. First, we performed the summation only
for the positive frequency part of cosine, but obviously other $2^d-1$
terms do contribute to $\Omega_M(\alpha)$ and can not be ignored.
Secondly, we are considering an Euclidean space-time manifold $M$,
so it is strange that $i$ appears in front of the action in the final form
(\ref{eq:PFP}).

	We may understand these points in the following way.
The TV invariant is a triangulation independent topological invariant.
If we regard it as a partition function, it must be the one
of some topological field theory, while the Einstein gravity is not.
But in 3-dimensions we have known for some time a topological gravity
model: the Chern-Simons-Witten (CSW) gravity.
Recall that the SU(2)$\times$SU(2)
Chern-Simons (CS) theory is on-shell equivalent to 3-dimensional Einstein
gravity with positive cosmological constant.
Everything goes well and is consistent if we regard $\Omega_M(\alpha)$
as a partition function of the CSW theory, rather than
one of the Einstein gravity. Indeed, the CS theory
has $i$ factor in front of the action in the path-integral,
and, moreover,
since the CS gravity is a first order formalism,
path-integration includes the sum over the orientation of space-time:
\beq
{
{\cal D}e{\cal D}\omega e^{i\int e\wedge R}={\cal D}|e|{\cal D}\omega\cdot
2\cos\int|e|\wedge R.
}
\eeq
So the appearance of cosine is naturally acceptable as the state-sum
over parity (Note that our measure ${\cal D\phi}$ (\ref{eq:dphi}) is positive
definite.).

	On the other hand, by comparing the representation
of the modular group induced from
the Jones polynomial with the one from the TV invariant,
it has been
conjectured in \cite{TV} that for any closed oriented
3-manifold $M$ the TV invariant
$\Omega_M$ is equal to the absolute square of Witten's
invariant \cite{WitJones} with corresponding level $k$.
This relation can be checked
in some simple topologies \cite{OS,KMS,RS} as is checked later.
Besides, as shown in appendix, the TV invariant satisfies
the factorization formula characteristic to CS theories, and hence
the check extends to arbitrary number of connected sums of
such topologies.\footnote{
Turaev has announced that this fact has been proved \cite{T}.
}
So we may write $\Omega_M$ as a partition function
of the CS theory with two independent SU(2) gauge group:
\beq
\Omega_M=\int{\cal D}A^+{\cal D}A^-e^{i(S_{CS}[A^+]-S_{CS}[A^-])}
\label{eq:CSW}
\eeq
\beq
S_{CS}[A]=\int\epsilon^{ijk}(2A_i^a\partial_jA_k^a
+\frac{2}{3}\epsilon_{abc}A_i^aA_j^bA_k^c).
\eeq
This is the CSW Euclidean gravity partition function
with positive cosmological constant. It agrees
with the next-leading estimation
for the asymptotic behavior of $q$-$6j$ symbols \cite{MT},
and the study of the physical Hilbert space of ISO(3) CS theory \cite{OS}.
Note that the relative sign between $S_{CS}[A^+]$
and $S_{CS}[A^-]$ comes from the fact that $\Omega_M$ is the absolute
value square, and hence the "exotic" term is absent in the CSW action
as it is in (\ref{eq:action}).
We thus identify the TV invariant with an exact Euclidean CSW partition
function and proceed further to see its relation to RCFT's.
\section{TV Invariant from RCFT's}
\setcounter{equation}{0}

\subsection{Axioms of RCFT}

\indent

We will now explain how we can construct the TV invariant
from RCFT. First, let us consider the dual cell subdivision of a tetrahedron.
The boundary of the dual surfaces
which lies on a 2-simplex of the tetrahedron
forms a 3-point vertex, and hence the tetrahedron may be regarded
as a matrix element which connects $s$- and $t$-channel amplitude (fig.3).
Indeed, this is the definitions of $6j$ symbol, where a vertex represents
a composition of two representations. Thus, in particular, one may consider
a tetrahedron as a fusion matrix of conformal blocks \cite{MS}.
One of the most important properties of 2-dimensional conformal
field theory is duality. In a RCFT the space of physical conformal blocks is
finite-dimensional, so that any two $N$-point blocks with the same
external lines are related by a sequence of fusion and braiding
represented by some finite-dimensional matrices. Such a sequence is not
unique in general, but the duality matrix can not depend on them
(after the phase which comes from the framing is specified).
Therefore the fusion and braiding matrices are required to satisfy
some polynomial equations. This is the idea of ref.\cite{MS} of an axiomatic
approach to RCFT. Representing a sequence of fusion by a triangulated
3-manifold $M$, where the boundaries of dual surfaces
lying on $F_{\pm}$ such that
$\partial M=F_+\cup F_-$ and $F_+\cap F_-=\phi$ represents two conformal
blocks, the fact that the duality matrix does not depend on the way of fusion
implies that it does not depend on the triangulation of $M$. Thus we may
expect the duality matrix of RCFT to be a 3-dimensional topological invariant.
We will show that this is the case.

	We will now summarize the axiom of RCFT \cite{MS}.
Let $I$ be a finite index set, each of which represents a primary field
of the chiral algebra ${\cal A}$,
with a distinguished element 0
that represents the identity operator.
$i^{\vee}$ is assumed to be the only field
that produces the identity operator {\bf 1}
by fusion with $i$.
Let $V^i_{jk}$ $(i,j,k\in I)$ be a space of chiral vertex operator
${\cal H}_j\otimes {\cal H}_k\rightarrow {\cal H}_i$, where ${\cal H}_i$
denotes the representation space of ${\cal A}$. A chiral vertex operator is
an intertwiner of representations of ${\cal A}$, {\it i.e.} an operator
such that commute with the action of ${\cal A}$.
$V^i_{jk}$ is assumed to be a finite dimensional: $\dim V^i_{jk}=N^i_{jk}$.
We represent an element of $V^i_{jk}$ by a trivalent vertex
as usual:
\beq
\raisebox{0mm}{
}
\eeq
We restrict ourselves to the case such that $N^i_{jk}$ is either 0 or 1,
so we can take the matrix representation for duality transformations.
The genus 0 duality transformations are generated
by the fusion $F$, the braiding $B(\pm)$ and the braiding
on a single chiral vertex $\Omega(\pm)$ (not to be confused
with $\Omega_M$ in the previous section), which are defined in the
following picture:
\beqa
\vspace{3mm}
\raisebox{-6mm}{
}
&=&\sum_qF_{pq}\left[
\begin{array}{cc}i&j\\k&l
\end{array}
\right]
\raisebox{-6mm}{
}\\
\vspace{3mm}
\raisebox{-6mm}{
}
&=&\sum_qB_{pq}\left[
\begin{array}{cc}i&j\\k&l
\end{array}
\right](+)
\raisebox{-6mm}{
}\\
\vspace{3mm}
\raisebox{-6mm}{
}
&=&\Omega^i_{jk}(+)
\raisebox{-6mm}{
}
\eeqa
and analogous relations for $B(-)$ and $\Omega(-)$.
Then the genus 0 equations are
\beqa
&&\sum_sF_{p_2s}\left[
\begin{array}{cc}j&k\\p_1&b
\end{array}
\right]F_{p_1l}\left[
\begin{array}{cc}i&s\\a&b
\end{array}
\right]F_{sr}\left[
\begin{array}{cc}i&j\\l&k
\end{array}
\right]
=F_{p_1r}\left[
\begin{array}{cc}i&j\\a&p_2
\end{array}
\right]F_{p_2l}\left[
\begin{array}{cc}r&k\\a&b
\end{array}
\right]
\label{eq:pentagon}
\\
&&\Omega_{lk}^m(\epsilon)
F_{mn}\left[
\begin{array}{cc}j&k\\i&l
\end{array}
\right]
\Omega_{jk}^n(\epsilon)
=\sum_r
F_{mr}\left[
\begin{array}{cc}j&l\\i&k
\end{array}
\right]
\Omega_{kr}^i(\epsilon)
F_{rn}\left[
\begin{array}{cc}k&j\\i&l
\end{array}
\right]
\;\;(\epsilon=\pm).
\eeqa
The braiding matrix $B$ is not independent
but is written in terms of $F$ and $\Omega$ as
\beq
B(\epsilon)=(\Omega(-\epsilon)\otimes{\bf 1})
F({\bf 1}\otimes\Omega(\epsilon)).
\label{eq:B=omega F omega}
\eeq
They satisfy
\beq
B(+)B(-)=B(-)B(+)={\bf 1}.
\label{eq:BB=1}
\eeq
For modular covariance we need three more constraints on the modular
$S(j)$ and $T$ matrices of the one-point function on the torus:
\beqa
&&S(j)^2=\pm Ce^{i\pi\Delta_j}\\
&&(S(j)T)^3=S(j)^2\\
&&(S\otimes{\bf 1})F({\bf 1}\otimes\Theta(-)\Theta(+))
	F^{-1}(S^{-1}\otimes{\bf 1})
=FPF^{-1}({\bf 1}\otimes\Omega(-)),\label{eq:SaS-1=b}
\eeqa
where $C$ is the conjugation operator,
$P$ is a flip operator which interchanges two chiral vertex operators,
and $\Theta(\pm)$ is defined by
\beq
\Theta(+)\left(\raisebox{-6mm}{
}\right)
=e^{+i\pi(\Delta_k-\Delta_i-\Delta_j)}
\raisebox{-6mm}{
}
\eeq
and an analogous relation for $\Theta(-)$.
Moore and Seiberg proved that once these conditions are satisfied,
then all the constraints that may arise from the requirement
of duality and modular covariance
at higher genus are guaranteed by them \cite{MS}.
They also showed that these general conditions that every RCFT must enjoy
are enough for the proof of Verlinde's conjecture \cite{V}.

	We would now like to show that the TV invariant in the last section
can be indeed constructed from these duality data.
First, the following relations can be easily checked:
\beqa
F_{pr}\left[
\begin{array}{cc}j&k\\i&l
\end{array}
\right]
&=&
\sigma_{13}\otimes\sigma_{23}
F_{p^{\vee}r}
\left[
\begin{array}{cc}k&j\\l^{\vee}&i^{\vee}
\end{array}
\right]
\sigma_{13}\otimes\sigma_{13}P\n
&=&
\sigma_{12}\otimes\sigma_{12}P
F_{pr^{\vee}}
\left[
\begin{array}{cc}i^{\vee}&l\\j^{\vee}&k
\end{array}
\right]
\sigma_{12}\otimes\sigma_{23}\n
&=&
\sigma_{123}\otimes\sigma_{132}P
F_{p^{\vee}r^{\vee}}\left[
\begin{array}{cc}l&i^{\vee}\\k^{\vee}&j
\end{array}
\right]
P
\sigma_{123}\otimes\sigma_{132},
\eeqa
where for instance, $\sigma_{23}:V^i_{jk}\rightarrow V^i_{kj}$ such that
\beq
\sigma_{23}(V^i_{jk})(\alpha\otimes\gamma\otimes\beta)
=V^i_{kj}(\alpha\otimes\beta\otimes\gamma)\;\;\;
(\alpha\in{\cal H}_{i^{\vee}},\;
\beta\in{\cal H}_{j},\;
\gamma\in{\cal H}_{k}),
\eeq
regarding $V^i_{jk}$ as a function on ${\cal H}_{i^{\vee}}\otimes{\cal H}_j
\otimes{\cal H}_k$.
We would like to identify $F$ as symbol of Turaev-Viro's initial data,
so we will consider a self-conjugate RCFT,
{\it i.e.} a theory in which $i$ is identical to $i^{\vee}$
for all fields $i\in I$.
Besides, if all the eigenvalues of $\sigma$ are $+1$,
we can take the orbit of $\sigma(V^i_{jk})$ as basis of
$(V^i_{jk},V^i_{kj},V^j_{ki},V^j_{ik},V^k_{ij},V^k_{ji})$
so that
\beq
F_{pr}\left[
\begin{array}{cc}j&k\\i&l
\end{array}
\right]
=
F_{pr}
\left[
\begin{array}{cc}k&j\\l&i
\end{array}
\right]
=
F_{pr}
\left[
\begin{array}{cc}i&l\\j&k
\end{array}
\right]
=
F_{pr}\left[
\begin{array}{cc}l&i\\k&j
\end{array}
\right].
\label{eq:Fsym}
\eeq
If some eigenvalues of $\sigma$ are $-1$,
(\ref{eq:Fsym}) holds only up to signs on such basis in general.
In the following we will consider for simplicity
a class of theories in which all the eigenvalues of $\sigma$
are $+1$. Such theories include the Virasoro minimal series.
We will soon comment on some other cases in which some eigenvalues are
$-1$.

	Now, in addition to (\ref{eq:Fsym}) one can prove
\beq
F_{nk}\left[
\begin{array}{cc}i&j\\l&p
\end{array}
\right]
F_{p0}\left[
\begin{array}{cc}k&k\\l&l
\end{array}
\right]
=
F_{pi}\left[
\begin{array}{cc}j&k\\n&l
\end{array}
\right]
F_{n0}\left[
\begin{array}{cc}i&i\\l&l
\end{array}
\right]
\label{eq:Fsym2}
\eeq
from the pentagon identity (\ref{eq:pentagon}).
Normalizing the 0th column of $F$ as
\beq
F_{k0}\left[
\begin{array}{cc}i&i\\j&j
\end{array}
\right]
=\sqrt{\frac{F_iF_j}{F_k}}\;\;(\mbox{\rm ``good gauge''}),
\eeq
where $F_k=F_{00}\left[
\begin{array}{cc}k&k\\k&k
\end{array}
\right]$, (\ref{eq:Fsym2}) reads
\beq
\sqrt{F_nF_k}
F_{nk}\left[
\begin{array}{cc}i&j\\l&p
\end{array}
\right]
=
\sqrt{F_pF_i}
F_{pi}\left[
\begin{array}{cc}j&k\\n&l
\end{array}
\right]\label{eq:Fsym'}
{}.
\eeq

\subsection{Construction of initial data}

\indent

Due to (\ref{eq:Fsym}) and (\ref{eq:Fsym'}) we may take
\beq
\left|\begin{array}{ccc}i&j&k\\l&m&n\end{array}\right|
=
\sqrt{F_nF_k}
F_{nk}\left[
\begin{array}{cc}i&j\\l&m
\end{array}
\right]
\label{eq:RCFTsymbol}
\eeq
as symbol with full tetrahedral symmetry.
Combining (\ref{eq:RCFTsymbol}) and (\ref{eq:pentagon}),
$(**)$ of (\ref{eq:initial data})
is satisfied if we take $w_j^2=F_j^{-1}$.
Furthermore, combining (\ref{eq:B=omega F omega}) and (\ref{eq:BB=1})
we find
\beq
F_{pq}\left[
\begin{array}{cc}j&l\\i&k
\end{array}
\right]
F_{qr}\left[
\begin{array}{cc}j&k\\i&l
\end{array}
\right]
=
\delta_{pr},
\eeq
which implies the condition $(*)$. Thus we have shown that
we can construct the TV invariant from duality data of a self-conjugate
RCFT with irreducible $I$. This is the case with the Virasoro minimal
series (and the SU(2) WZW model), since successive fusions of the ``shift''
operators \cite{BPZ}, like $\phi_{1,2}$ and $\phi_{2,1}$
operators in the former theory, connect any two primary fields in them.

	In the SU(2) WZW model it is well-known that the duality
matrices are written by using the
${\cal U}_q$su(2) $q$-$6j$ symbol, and in particular \cite{AGS}
\beq
F_{kn}\left[
\begin{array}{cc}i&m\\j&l
\end{array}
\right]
=\sqrt{[2k+1][2n+1]}(-1)^{i+j+l+m}
\left\{
\begin{array}{ccc}
i&j&k \\ l&m&n
\end{array}
\right\}.
\eeq
An eigenvalue of $\sigma$ can be $\pm 1$ in this model.
For example, the eigenvalue of $\sigma_{23}$ on $V^i_{jj}$
is $+(-)1$ if the representation $i$ occurs (anti-)symmetrically
in the tensor product ${\cal H}_j\otimes{\cal H}_j$.
However, one can take appropriate basis of the chiral vertex operators
(Racah-Wigner normalization) to obtain a symmetric symbol
\beq
\left|\begin{array}{ccc}i&j&k\\l&m&n\end{array}\right|
=(-1)^{i+j+k+l+m+n}\left\{\begin{array}{ccc}i&j&k\\l&m&n\end{array}\right\}.
\eeq
Hence the original Turaev-Viro's initial data
(\ref{eq:su(2) initial data}) is associated with the $A_{k+1}$
modular invariant SU(2) WZW model.
Although we have restricted ourselves to the simplest RCFT's, we may
do the same thing in other self-conjugate RCFT's\footnote{
For non-self-conjugate theories
Durhuus {\it et. al.} have constructed a generalized
topological invariant to such cases \cite{DJN}, in which symbols
possess only the half of the tetrahedral symmetry that preserves
the orientation.}
with left right
factorized $N^i_{jk}$, such as the $D_{2\rho+2}$ with $\rho$ even,
$E_6$ and $E_8$ modular invariant SU(2) WZW model \cite{ICZ,BYZ}.
The latter two are easy because $N^i_{jk}$ is either 0 or 1 also in these
models. We will list the initial data constructed from those:\\
$\cdot E_6$ case
\newcommand{\1}{\bf 1}
\beqa
\mbox{\rm fusion rule}&:&\psi\times\psi=\1\n
&&\psi\times\sigma=\sigma\n
&&\sigma\times\sigma=\1+\psi\nonumber
\eeqa
\beqa
&\left|\begin{array}{ccc}\1&\1&\1\\ \1&\1&\1 \end{array}\right|=1~~~~
&\left|\begin{array}{ccc}\1&\psi&\psi\\
\1&\psi&\psi\end{array}\right|=1\n
&&\n
&\left|\begin{array}{ccc}\1&\1&\1\\ \psi&\psi&\psi \end{array}\right|=1~~~
&\left|\begin{array}{ccc}\1&\1&\1\\
\sigma&\sigma&\sigma\end{array}\right|
=2^{-\frac{1}{4}}\n
&&\n
&\left|\begin{array}{ccc}\sigma&\sigma&\1\\
\psi&\psi&\sigma\end{array}\right|
=2^{-\frac{1}{4}}
&\left|\begin{array}{ccc}\sigma&\sigma&\1\\
\sigma&\sigma&\1\end{array}\right|
=2^{-\frac{1}{2}}\n
&&\n
&\left|\begin{array}{ccc}\sigma&\sigma&\1\\
\sigma&\sigma&\psi\end{array}\right|
=2^{-\frac{1}{2}}
&\left|\begin{array}{ccc}\sigma&\sigma&\psi\\
\sigma&\sigma&\psi\end{array}\right|
=-2^{-\frac{1}{2}}\nonumber
\eeqa
\beqa
&w_{\1}^2=1~~~&w_{\sigma}^2=2^{\frac{1}{2}}\n
&w_{\psi}^2=1~~~&w^2=\tilde{w}^2=4 \label{eq:E6}
\eeqa
$\cdot E_8$ case
\beqa
\mbox{\rm fusion rule}&:&\varphi\times\varphi=\1+\varphi\nonumber
\eeqa
\beqa
&\left|\begin{array}{ccc}\1&\1&\1\\ \1&\1&\1 \end{array}\right|=1~~~~~~~
&\left|\begin{array}{ccc}\1&\varphi&\varphi\\
\1&\varphi&\varphi\end{array}\right|=[2]^{-1}\n
&&\n
&\left|\begin{array}{ccc}\1&\varphi&\varphi\\
\varphi&\1&\1\end{array}\right|=[2]^{-1}~~
&\left|\begin{array}{ccc}\1&\varphi&\varphi\\
\varphi&\varphi&\varphi\end{array}\right|=-[2]^{-1}\n
&&\n
&\left|\begin{array}{ccc}\varphi&\varphi&\varphi\\
\varphi&\varphi&\varphi\end{array}\right|=-[2]^{-2}&\nonumber
\eeqa
\beqa
&w_{\1}^2=1~~~&w_{\varphi}^2=[2]\n
&w^2=\tilde{w}^2=1+[2]^2~~~&[2]=\frac{\sin\frac{2\pi}{5}}{\sin\frac{\pi}{5}},
\label{eq:E8}
\eeqa
where the primary fields are the ones of WZW models with extended
chiral algebra, {\it i.e.} level-1 $C_2$ ($E_6$ case) and
level-1 $G_2$ ($E_8$ case), respectively \cite{E6,E8}.
The fusion rule for the $E_6$ case are the same as the Ising one,
and the initial data (\ref{eq:E6}) are easily read off
from the appendix of ref.\cite{MS}.
The fusion rule for the $E_8$ case are obtained from
level-3 SU(2) WZW model by restricting primary fields
to integer-spin ones, and hence the initial data (\ref{eq:E8})
are given by $q$-$6j$ symbols.

	It is useful to relate $\tilde{w}$ with $00$ entry
of the modular $S$ matrix. It is known that the condition (\ref{eq:SaS-1=b})
can be used to solve for $S$ in terms of $F$ and $B$,
and in particular
\beq
S_{i0}=S_{00}F_i^{-1}
\eeq
in this gauge. $S_{00}$ can be fixed by unitarity of $S$:
\beq
S_{00}^2=\tilde{w}^{-2}.
\eeq
Note that
\beqa
1&=&\sum_lw_l^2
\left|\begin{array}{ccc}i&k&l\\k&i&0\end{array}\right|
\left|\begin{array}{ccc}k&k&0\\i&i&l\end{array}\right|\n
&=&\sum_lw_l^2w_i^{-2}w_k^{-2}\delta(i,k,l)\n
&=&c(i,k),
\eeqa
and hence
\beq
\tilde{w}^2=w^2=S_{00}^{-2}. \label{eq:w=1/S00}
\eeq
This relation is known in the case associated with
the $A_{k+1}$ modular invariant SU(2) WZW model (the original Turaev-Viro's
initial data). It can be easy to check that (\ref{eq:w=1/S00})
holds true also for the $E_6$ and the $E_8$ case.
We would like to stress here that it
is a direct consequence of the requirement of modular covariance of RCFT.

\subsection{Examples}

\indent

	We will	now calculate the Turaev-Viro invariant for $S^3$
associated with duality data of a RCFT. It would be most easily done
by presenting an $S^3$ as two 3-ball $B^3$ whose boundaries are glued
together, but to illustrate the idea for finding upper bound
in the next section
we will present it here as two solid tori such that the boundary
of the one is identified with that of the other after the modular $S$
transformation.

	Since the TV invariant does not depend on the triangulation
within the manifold, we may take any triangulation of the solid torus.
A convenient choice is such that its boundary, which is a torus,
is triangulated as in fig.4 with opposite sides of the rectangle identified.
Its dual graph represents a genus 3 conformal block.
It is straightforward to see that for such coloring
$\alpha = \alpha(j,l,i,k,i',k')$:
\beq
\Omega_{D^2\times S^1}(\alpha(j,l,i,k,i',k'))
=\frac{w_jw_l}{w^2}\delta_{ii'}\delta_{kk'}.
\label{eq:D2S2}
\eeq
The modular $S$ transformation on $\alpha(j,l,i,k,i',k')$
is performed by interchanging $a$- (meridian)
and $b$- (longitude) cycle
($a\mapsto b$, $b\mapsto -a$), and it maps the genus 3 conformal block
to another one. These two blocks are related
by fusing twice (fig.5), and hence
\beqa
&&\Omega_{D^2\times S^1}(S(\alpha(j,l,i,k,i',k')))\n
&=&\sum_{j',l'}
F_{ll'}\left[
\begin{array}{cc}i&k\\i'&k'
\end{array}
\right]
F_{jj'}\left[
\begin{array}{cc}i&k\\i'&k'
\end{array}
\right]
\Omega_{D^2\times S^1}(\alpha(l',j',i,i',k,k')).
\label{eq:D2S2'}
\eeqa
It may be noted that in the case of the Jones polynomial
the modular transformation is represented by the modular $S$ matrix,
while in our case it is represented by the duality matrices.
In contrast to the former case there is no phase ambiguity
since it depends only on the coloring of the boundary \cite{TV}.
It is also noticed that the $F$ transformation induces the Alexander
move on the triangulated 2-surface.
Combining (\ref{eq:D2S2}) and (\ref{eq:D2S2'}) for $S^3$
\beqa
\Omega_{S^3}&=&\sum_{j,l,i,k,i',k'}
\Omega_{D^2\times S^1}(\alpha(j,l,i,k,i',k'))
\Omega_{D^2\times S^1}(S(\alpha(j,l,i,k,i',k')))\n
&=&\sum_{j,l,i,k,i',k'}
\sum_{j',l'}
F_{ll'}\left[
\begin{array}{cc}i&k\\i'&k'
\end{array}
\right]
F_{jj'}\left[
\begin{array}{cc}i&k\\i'&k'
\end{array}
\right]
\frac{w_{l'}w_{j'}}{w^2}\delta_{ik}\delta_{i'k'}
\frac{w_jw_l}{w^2}\delta_{ii'}\delta_{kk'}
\n
&=&\sum_{j,l,i}
\sum_{j',l'}
F_{ll'}\left[
\begin{array}{cc}i&i\\i&i
\end{array}
\right]
F_{jj'}\left[
\begin{array}{cc}i&i\\i&i
\end{array}
\right]
\frac{w_{l'}w_{j'}}{w^2}
\frac{w_jw_l}{w^2}
\n
&=&\sum_{j,l,i}
\sum_{j',l'}
F_{ll'}\left[
\begin{array}{cc}i&i\\i&i
\end{array}
\right]
F_{l'0}\left[
\begin{array}{cc}i&i\\i&i
\end{array}
\right]
F_{jj'}\left[
\begin{array}{cc}i&i\\i&i
\end{array}
\right]
F_{j'0}\left[
\begin{array}{cc}i&i\\i&i
\end{array}
\right]
\frac{w_i^4w_jw_l}{w^4}
\n
&=&
\sum_{j,l,i}
\delta_{l0}\delta_{j0}
\frac{w_i^4w_jw_l}{w^4}
\n
&=&\frac{\tilde{w}^2}{w^4}\n
&=&w^{-2}\n
&=&S_{00}^2
{}.
\label{eq:omegaS3}
\eeqa
Here in the first line
we do not need to reverse one of $\alpha(j,l,i,k,i',k')$
because the TV invariant is independent of the orientation of the manifold.
This result is also well-known in the case of the original Turaev-Viro's
initial data, and is consistent with the fact
that the TV invariant is an absolute value
square of Witten's invariant.
For the same reason, the TV invariant associated with the $E_6$
and the $E_8$ modular invariant may be identified as the
${\rm SO(5)}\times{\rm SO(5)}$
and the $G_2\times G_2$ CS partition function, respectively.

	We can calculate also for $S^2\times S^1$
in the same way and find
\beqa
\Omega_{S^2\times S^1}&=&\sum_{j,l,i,k,i',k'}
\left(\Omega_{D^2\times S^1}(\alpha(j,l,i,k,i',k'))\right)^2
\n
&=&1.
\eeqa
Hence obviously $\Omega_{S^3}$ is always smaller
than $\Omega_{S^2\times S^1}$.
In the next section we will generalize this fact to multi-wormhole
partition functions.

\section{Multi-wormhole partition function}
\setcounter{equation}{0}

\indent

	We will now go back to the case of the original Turaev-Viro's
initial data, which is relevant to the CSW gravity.
It is known that any closed, orientable 3-manifold $M$
is presented by so-called ``Hegaard splitting'' \cite{3-manifolds}.
We say a 3-manifold $M$ admits a Hegaard splitting of genus $g$
if $M$ is obtained by identifying boundaries of a pair of genus $g$
handlebodies $(M_1,M_2)$ through a modular transformation $\varphi$,
{\it i.e.}
\beq
M=M_1\cup_{\varphi} M_2, \;\;\partial M_1\sim -\partial M_2\sim \Sigma_g.
\eeq
Here $\Sigma_g$ denotes a genus $g$ 2-surface.
$\partial M_1$ is identified with $\varphi(\partial M_2)$
after reversing the orientation.
Regarding $\Omega_M$ as the CSW gravity partition function $Z(M)$,
we can in principle calculate $Z(M)$ for any closed,
orientable 3-manifold $M$.

	We first consider a genus $g$ Hegaard splitting of $g$ times
connected sum $(S^2\times S^1)\sharp\cdots\sharp (S^2\times S^1)$
of $S^2\times S^1$. Here a 3-manifold $M$ is said to be a connected sum
$M_1\sharp M_2$ of $M_1$ and $M_2$ if $M$ is obtained by cutting
3-balls from each of $M_1$ and $M_2$,
and then identifying the resulting boundaries.
One can obtain $(S^2\times S^1)^{\sharp g}$
by taking $\varphi={\mbox{\rm identity}}$
and gluing two genus $g$ handlebodies together. So
\beqa
Z((S^2\times S^1)^{\sharp g})&\equiv&\Omega_{(S^2\times S^1)^{\sharp g}}\n
&=&\sum_{\alpha}(\Omega_{H_g}(\alpha))^2, \label{eq:inner product}
\eeqa
where $H_g$ denotes the handlebody of genus $g$, and $\alpha$ is a coloring
of some triangulation of $\Sigma_g\sim\partial H_g$.
On the other hand we can calculate $\Omega_{(S^2\times S^1)^{\sharp g}}$
by using the factorization formula:
\beq
\frac{\Omega_{M_1\sharp M_2}}{\Omega_{S^3}}
=
\frac{\Omega_{M_1}}{\Omega_{S^3}}
\frac{\Omega_{M_1}}{\Omega_{S^3}},
\label{eq:factor}
\eeq
which we will prove in appendix.\footnote{
(\ref{eq:factor}) has been proved also in ref.(\cite{KMS})
by modifying the construction of invariants for 3-manifolds with boundary.
}
(\ref{eq:factor}) is a characteristic property of the partition function
of the CS theory \cite{WitJones}; the fact that the TV
invariant satisfies (\ref{eq:factor}) is a reflection
of the equivalence to the CSW theory. Hence
\beqa
Z((S^2\times S^1)^{\sharp g})
&=&
\left(\frac{\Omega_{S^2\times S^1}}{\Omega_{S^3}}\right)^g\cdot \Omega_{S^3}
\n
&=&(\Omega_{S^3})^{-g+1}\n
&=&\left(-\frac{(q-q^{-1})^2}{2(k+2)}\right)^{-g+1},
\eeqa
where
\beqa
\Omega_{S^3}&=&S_{00}^2\n
&=&-\frac{(q-q^{-1})^2}{2(k+2)}\;\;\;(q=e^{\frac{i\pi}{k+2}})
\eeqa
has been used. Note that in the large $k$ limit
$Z(S^3)\equiv\Omega_{S^3}$ behaves as
\beq
Z(S^3)\sim k^{-3}\sim \Lambda^{\frac{3}{2}}.
\eeq

	Now, how about other topologies? Suppose that $M$ admits
a genus $g$ Hegaard splitting with gluing transformation $\varphi$.
We write
\beq
Z(M)=\sum_{\alpha}\Omega_{H_g}(\alpha)\Omega_{H_g}(\varphi(\alpha)).
\eeq
In the previous section we saw that the modular
transformation on the torus was described by the duality matrices between
two genus 3 conformal blocks represented by the dual graphs of
triangulated tori. In general the modular group is generated
by the Dehn twists along non-contractible
homology 1-cycles, under which a conformal block,
represented by the dual graph of a triangulated boundary,
obviously does not change its genus. Hence we know that
these two conformal blocks are related
by the duality matrices uniquely,
at least up to phase which comes from the framing.
But in our case there is no such phase ambiguity
because the TV invariant is real except boundary factors, which are fixed.
Thus the modular transformation is described by the duality matrices
between the conformal blocks associated with the dual graphs.
Since the duality matrices are unitary, we use
the Cauchy-Schwartz inequality to obtain
\beq
Z(M)\leq Z((S^2\times S^1)^{\sharp g}).
\eeq
Hence the partition function $Z(M)$ is bounded
by $Z((S^2\times S^1)^{\sharp g})$,
where $g$ is the smallest genus in which $M$ can be presented
by Hegaard splitting.
This is in agreement with the estimation for Witten's invariant
by Kohno \cite{Kohno} together with the fact that the TV invariant is
its absolute square. Since
\beq
Z((S^2\times S^1)^{\sharp g})
\sim\Lambda^{-\frac{3g-3}{2}} \label{eq:upper bound}
\eeq
in the large $k$, this upper bound diverges as $\Lambda\rightarrow 0$
if $g\geq 2$.
However, the ratio
$\frac{Z(M)}{Z((S^2\times S^1)^{\sharp g})}$ is generically $O(1)$
unless the ``angle''
between $\Omega_{H_g}(\alpha)$ and $\Omega_{H_g}(\varphi(\alpha))$
is accidentally very near $\frac{\pi}{2}$.
Therefore in the CSW gravity we may say that
$Z(M)$ is generically very large if $M$ is neither $S^3$ nor a lens
space, and many-wormhole
configurations dominate near $\Lambda\sim +0$ in the sense
that $Z(M)$ generically tends to
diverge faster as the ``number of
wormholes'' $g$ becomes larger.
However, in summation over wormholes one should take their statistics
into account, and hence $g+1$ wormhole partition function is suppressed
by $((g+1)!)^{-1}$ if one
assumes that wormholes have bosonic statistics.
Besides, it has been argued \cite{CD} that one should consider
the slide diffeomorphism of 3-manifolds as a part of gauge group,
and then a contribution from a many-wormhole configuration
would be more suppressed in the wormhole summation.

\section{Conclusion}

\indent

	In this paper we have studied the TV invariant
as the partition function of the Euclidean CSW gravity with positive
cosmological constant. We have shown that initial data of the TV invariant
can be
constructed from the duality matrices of a self-conjugate RCFT with
symmetrizable fusion matrix, and that in particular the original
Turaev-Viro's initial data is associated with those of the $A_{k+1}$ modular
invariant SU(2) WZW model.
The partition function $Z(M)$ has been shown to be bounded from above
by $Z((S^2\times S^1)^{\sharp g})
=(S_{00})^{-2g+2}\sim \Lambda^{-\frac{3g-3}{2}}$,
where $g$ is the smallest genus of handlebodies with which $M$
can be presented by Hegaard splitting.
$Z(M)$ is generically very large near $\Lambda\sim +0$
if $M$ is neither $S^3$ nor a lens
space, and many-wormhole
configurations dominate near $\Lambda\sim +0$ in the sense
that $Z(M)$ generically tends to
diverge faster as the ``number of
wormholes'' $g$ becomes larger.

	The fact that the value of the TV invariant on $S^3$
is always $S_{00}^2$ is a direct consequence of modular covariance
of RCFT, though it is still obscure intrinsically why the topological
invariant constructed from the duality matrices in such a way gives
the absolute square of Witten type topological invariant.
The TV invariant may be considered as a lattice realization
of the CS partition theory,
so it would be also interesting to ask
whether it can be generalized to $2+1$-dimensional gravity.
In that case we may need to study the duality matrix of
WZW theories with some non-compact gauge groups,
in which we do not know any distinguished finite set of
representations such as ``good'' representations in the ${\cal U}_q$sl(2).
In that sense it is not clear how to do so
because rationality of underlying CFT
is essential in regularization for the divergence in the state-sum
on a lattice.

\section*{Acknowledgement}

\indent

	The author is grateful to M.Ninomiya for useful discussions
and comments, and also to M.Hayashi for discussions on 3-manifolds.

\section*{Appendix}

\indent

	In this appendix we will give an elementary proof of the factorization
formula (\ref{eq:factor}):
\renewcommand{\theequation}{\mbox{4.3}}
\beq
\frac{\Omega_{M_1\sharp M_2}}{\Omega_{S^3}}
=
\frac{\Omega_{M_1}}{\Omega_{S^3}}
\frac{\Omega_{M_2}}{\Omega_{S^3}}
\nonumber
\eeq
\setcounter{equation}{0}
\renewcommand{\theequation}{\mbox{A.\arabic{equation}}}
Consider first a triangulated cylinder $D^2\times[0,1]$
as shown in fig.6. Such coloring of the triangulation of
its boundary is denoted by $\beta$.
It is made of three tetrahedra, so
\beqa
\Omega_{D^2\times[0,1]}(\beta)
&=&
\left|\begin{array}{ccc}i&j&k\\l&m&n\end{array}\right|
\left|\begin{array}{ccc}p&m&i'\\l&q&k\end{array}\right|
\left|\begin{array}{ccc}i'&j'&k'\\r&q&l\end{array}\right|\n
&&\cdot w_iw_jw_kw_{i'}w_{j'}w_{k'}
w_lw_mw_nw_pw_qw_rw^{-6}.
\eeqa
Gluing two such cylinders together, we have
\beqa
&&\Omega_{S^2\times[0,1]}(\gamma(i,j,k),\gamma(i',j',k'))\n
&=&\sum_{\beta}(\Omega_{D^2\times[0,1]}(\beta))^2
(w_iw_jw_kw_{i'}w_{j'}w_{k'}w^{-6})^{-1}\n
&=&\sum_{l,m,n,p,q,r}
\left(
\left|\begin{array}{ccc}i&j&k\\l&m&n\end{array}\right|
\left|\begin{array}{ccc}p&m&i'\\l&q&k\end{array}\right|
\left|\begin{array}{ccc}i'&j'&k'\\r&q&l\end{array}\right|
w_lw_mw_nw_pw_qw_r
\right)^2\n
&&\cdot w_iw_jw_kw_{i'}w_{j'}w_{k'}w^{-6}\n
&=&\sum_{l,m,q}
\left(
\sum_n
\left|\begin{array}{ccc}i&j&k\\l&m&n\end{array}\right|^2
w_k^2w_n^2
\right)
\left(
\sum_p
\left|\begin{array}{ccc}p&m&i'\\l&q&k\end{array}\right|^2
w_k^2w_n^2
\right)
\left(
\sum_r
\left|\begin{array}{ccc}i'&j'&k'\\r&q&l\end{array}\right|^2
w_k^2w_n^2
\right)\n
&&
\cdot w_iw_jw_k^{-1}w_{i'}^{-1}w_{j'}w_{k'}w_m^2w_q^2w^{-6}\n
&=&
\sum_{l,m,q}
w_iw_jw_k^{-1}w_{i'}^{-1}w_{j'}w_{k'}w_m^2w_q^2w^{-6}
\delta(i',q,l)\delta(l,m,k)\n
&=&
\sum_l
w_iw_jw_kw_{i'}w_{j'}w_{k'}w_l^4w^{-6}c^2\n
&=&
w_iw_jw_kw_{i'}w_{j'}w_{k'}w^{-4}c
\eeqa
where
$\gamma(i,j,k)$ represents the coloring of the triangulation
of $S^2$ (fig.7).
On the other hand, we can calculate $\Omega_{B^3}(\gamma(i,j,k))$
by gluing two tetrahedra (fig.8) to obtain
\beqa
\Omega_{B^3}(\gamma(i,j,k))
&=&\sum_{l,m,n}
\left|\begin{array}{ccc}i&j&k\\l&m&n\end{array}\right|^2
w_iw_jw_kw_l^2w_m^2w_n^2w^{-5}\n
&=&w_iw_jw_kw^{-3}.
\eeqa
Therefore
\beq
\Omega_{S^2\times[0,1]}(\gamma(i,j,k),\gamma(i',j',k'))
=
\Omega_{B^3}(\gamma(i,j,k))
\Omega_{B^3}(\gamma(i',j',k'))\cdot
w^2c.
\eeq
Using $\Omega_{S^3}=(w^2c)^{-1}$ (\ref{eq:omegaS3}), we have
\beq
\frac{\Omega_{S^2\times[0,1]}(\gamma(i,j,k),\gamma(i',j',k'))}
{\Omega_{S^3}}
=
\frac{\Omega_{B^3}(\gamma(i,j,k))}
{\Omega_{S^3}}
\frac{\Omega_{B^3}(\gamma(i',j',k'))}
{\Omega_{S^3}}.
\label{eq:connect}
\eeq
(\ref{eq:factor}) immediately follows from (\ref{eq:connect}).

\newpage
\begin{figure}[p]
\caption{Subdivision dual to triangulation.}
\vskip 10mm
\centerline{
\raisebox{0mm}{
}}
\end{figure}
\begin{figure}[p]
\caption{An Alexander move
can be generated by compositions of
(a)the bubble move ${\cal B}$, (b)the lune move ${\cal L}$
and (c)the Matveev move ${\cal M}$. }
\vskip 5mm
\centerline{
\raisebox{0mm}{
}}
\centerline{
\raisebox{0mm}{
}}
\centerline{
\raisebox{0mm}{
}
}
\end{figure}
\begin{figure}[h]
\caption{A tetrahedron can be seen as a fusion matrix.}
\vskip 10mm
\centerline{
\raisebox{0mm}{
}
}
\end{figure}
\begin{figure}[h]
\caption{A convenient triangulation of a torus.
Opposite sides of the rectangle are identified.
Its dual graph represents a genus 3 conformal block.}
\vskip 10mm
\centerline{
\raisebox{0mm}{
}
}
\end{figure}
\begin{figure}
\caption{The two genus 3 conformal blocks are related
by fusing twice.}
\vskip 10mm
\centerline{
\raisebox{0mm}{
}
}
\end{figure}
\begin{figure}
\caption{A triangulated cylinder $D^2\times[0,1]$.}
\vskip 10mm
\centerline{
\raisebox{0mm}{
}}
\end{figure}
\begin{figure}
\caption{$S^2\times[0,1]$ obtained by gluing two cylinders
together.}
\vskip 10mm
\centerline{
\raisebox{0mm}{
}}
\end{figure}
\begin{figure}
\caption{$B^3$ obtained by gluing two tetrahedra.}\vskip 10mm
\centerline{
\raisebox{0mm}{
}}
\end{figure}
\end{document}